\documentclass[preprint2]{aastex}

\shorttitle{\indent \def Observations of Sunspot Bright Dots with IRIS} \shortauthors{Tian et al.}

\begin{document}

\title{Observations of Subarcsecond Bright Dots in the Transition Region above Sunspots with the Interface Region Imaging Spectrograph}

\author{H. Tian\altaffilmark{1}, L. Kleint\altaffilmark{2}, H. Peter\altaffilmark{3}, M. Weber\altaffilmark{1}, P. Testa\altaffilmark{1},  E. DeLuca\altaffilmark{1}, L. Golub\altaffilmark{1},  N. Schanche\altaffilmark{1}}

\altaffiltext{1}{Harvard-Smithsonian Center for Astrophysics, 60 Garden Street, Cambridge, MA 02138; hui.tian@cfa.harvard.edu}
\altaffiltext{2}{University of Applied Sciences and Arts Northwestern Switzerland, Bahnhofstr. 6, 5210 Windisch, Switzerland}
\altaffiltext{3}{Max Planck Institute for Solar System Research, 37077 G\"ottingen, Germany}

\begin{abstract}
Observations with the Interface Region Imaging Spectrograph (IRIS) have revealed numerous sub-arcsecond bright dots in the transition region above sunspots.
These bright dots are seen in the 1400\AA{} and 1330\AA{} slit-jaw images. They are clearly present in all sunspots we investigated, mostly in
the penumbrae, but also occasionally in some umbrae and light bridges. The bright dots in the penumbrae typically appear slightly elongated, 
with the two dimensions being 300--600 km and 250--450 km, respectively. The long sides of these dots are often nearly parallel to the bright filamentary structures in the penumbrae but sometimes clearly deviate from the radial direction. Their lifetimes are mostly less than one minute, although some dots last for a few minutes or even longer. Their intensities are often a few times
stronger than the intensities of the surrounding environment in the slit-jaw images. About half of the bright dots show apparent
movement with speeds of $\sim$10--40~km~s$^{-1}$ in the radial direction. Spectra of a few bright dots were obtained and the
Si~{\sc{iv}}~1402.77\AA{} line profiles in these dots are significantly broadened. The line intensity can be enhanced by one to two orders of
magnitude. Some relatively bright and long-lasting dots are also observed in several passbands of the Atmospheric Imaging Assembly 
onboard the Solar Dynamics Observatory, and they appear to be located at the bases of loop-like structures. Many of these
bright dots are likely associated with small-scale energy release events at the transition region footpoints of magnetic loops.
\end{abstract}


\section{Introduction}

Sunspots are regions with the strongest magnetic field in the solar atmosphere. In the layers of photosphere and chromosphere, a sunspot usually has a
dark umbra and a less dark penumbra enclosing the umbra. The penumbra consists of filamentary structures which often have
dark cores \citep{Scharmer2002}. In photospheric images, grain-like features at the head of penumbral filaments are often found to move with speeds of a
few hundred meters per second \citep{Muller1973}. 
For reviews of sunspots, we refer to \cite{Solanki2003}, \cite{Borrero2011}, and \cite{Rempel2011}.

The solar transition region (TR) is the interface between the chromosphere and corona, where the temperature and density change dramatically.
Although sunspots have been extensively studied for centuries, the TR above sunspots is still poorly understood due to the lack of
high-resolution observations. Most existing investigations of the sunspot's TR are focused on oscillations
\citep[e.g.,][]{Brynildsen1999a,Brynildsen1999b,OShea2002,Tian2014}, supersonic downflows \citep{Brynildsen2001}, and sunspot plumes
\citep[e.g.,][]{Foukal1974,Doyle2003,Brosius2005,Tian2009}. 

The recently launched Interface Region Imaging Spectrograph \citep[IRIS,][]{DePontieu2014} mission is now providing high-resolution ($\sim$0.33$^{\prime\prime}$) observations of the TR and chromosphere. With IRIS observations, here
we report a new feature in the sunspot's interface region: subarcsecond bright dots. We measure their physical properties, 
investigate their spectroscopic signatures and relate them to coronal structures seen in images taken by the Atmospheric Imaging Assembly
\citep[AIA,][]{Lemen2012} onboard the Solar Dynamics Observatory \citep[SDO,][]{Pesnell2012}.  

\section{Data analysis}

\begin{figure*}
\centering {\includegraphics[width=\textwidth]{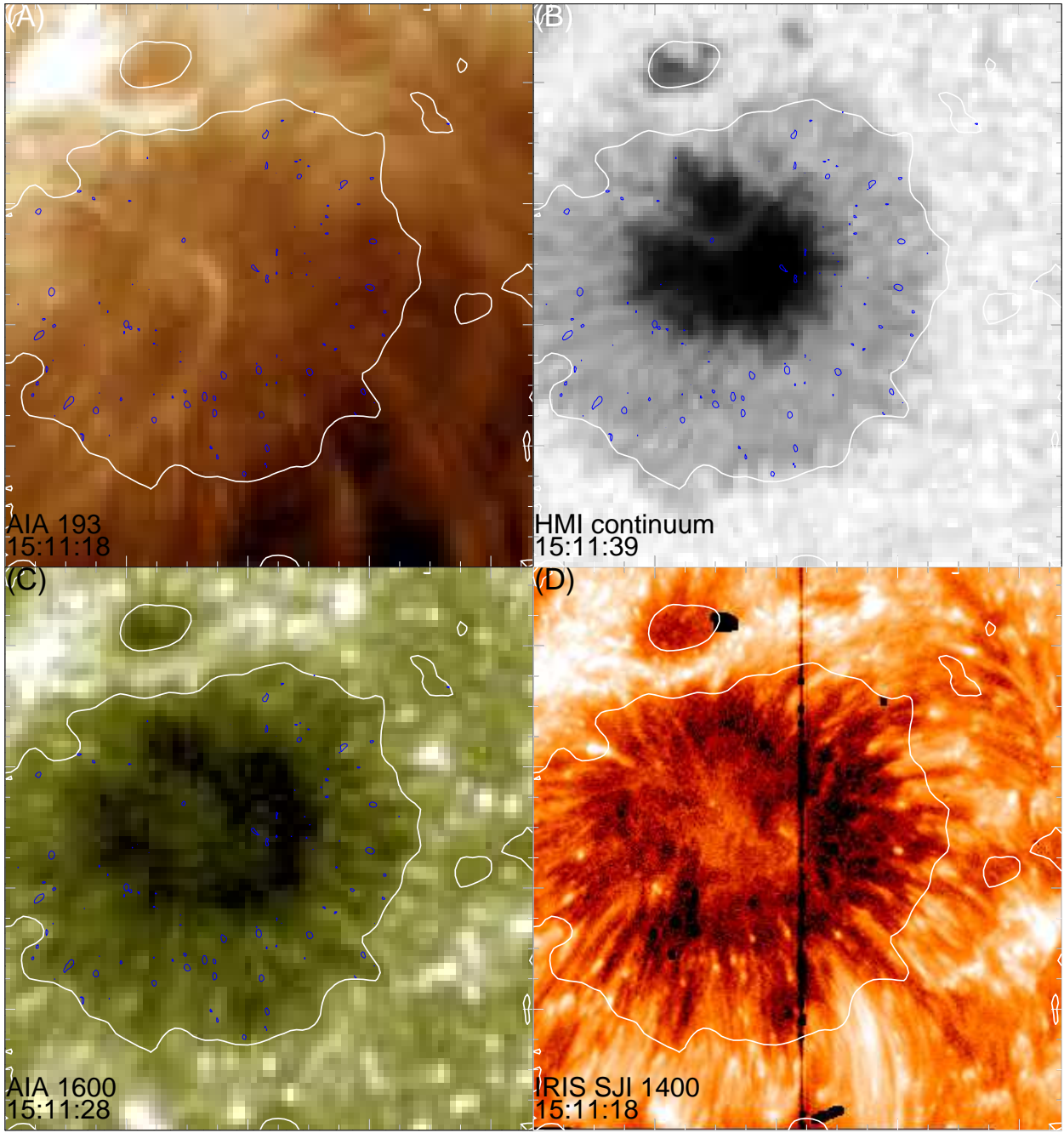}} \caption{ Images obtained with SDO and IRIS around 15:11 UT on 2014 January 9. The field of view has a size of
54$^{\prime\prime}$$\times$58$^{\prime\prime}$. The white contours derived from AIA
1600\AA{} intensity outline the spatial extension of the sunspot. The blue contours mark locations of bright dots within the sunspot. Two online
movies (m1.mpg, m2.mpg) are associated with this figure. } \label{fig.1}
\end{figure*}

We mainly analyze the sit-and-stare observation made from 14:39 to 15:29 on 2014 January 9. The pointing was (351$^{\prime\prime}$,
-127$^{\prime\prime}$), targeting the smaller sunspot of AR 11944. One thousand slit-jaw images (SJI) in the 1400\AA{} filter were obtained. Several spectral windows covering a few strong emission lines (e.g., Si~{\sc{iv}}~1402.77\AA{}, C~{\sc{ii}}~1334.53\AA{}
and Mg~{\sc{ii}}~K~2796.35\AA{}) were recorded. The cadence, exposure time and spatial pixel size are 3 seconds, 2 seconds and 0.167$^{\prime\prime}$, respectively. Dark current subtraction, flat field and geometrical corrections have been applied in the level 2 data used here.

This dataset was taken during the eclipse season. To remove the significant jitter and drift we have aligned the slit-jaw images to the first frame of every 150 images through cross correlation. 

Figure~\ref{fig.1}(A)-(C) shows the 193\AA{} and 1600\AA{} images taken by AIA and the continuum image taken by the Helioseismic and Magnetic Imager \citep[HMI,][]{Scherrer2012} onboard SDO. Cross-correlation between the AIA 1600\AA{} and IRIS 1400\AA{}~images are used for the coalignment, with an uncertainty of about one AIA pixel (0.6$^{\prime\prime}$). Some filamentary structures in the IRIS 1400\AA{}~image (Figure~\ref{fig.1}(D)) seem to be emission of the UV continuum (formed slightly above temperature minimum) and associated with the penumbral filaments in the HMI continuum image. Others are likely Si~{\sc{iv}}~emission ($\sim$10$^{4.9}$K) from legs of corona loops seen in AIA 193\AA{}.

We find many transient grain-like bright dots in the 1400\AA{} images (Figure~\ref{fig.1}(D), movie m1.mpg). They are mostly present in the penumbra, but also occasionally appear in the umbra and light bridge.
Some of them are clearly moving either inward or outward in the radial direction of the sunspot. We smooth each image over an area of
8$\times$8 pixels and subtract this smoothed image from the original image. The small-scale bright dots in the sunspot can be outlined using
contours of the resulting image (blue contours in Figure~\ref{fig.1}). We do not find obvious signatures of these dots in the HMI continuum image. A comparison between the IRIS and AIA images (movie m2.mpg) suggests that
some bright dots are associated with faint loops in AIA 193\AA{}.

\begin{figure*}
\centering {\includegraphics[width=\textwidth]{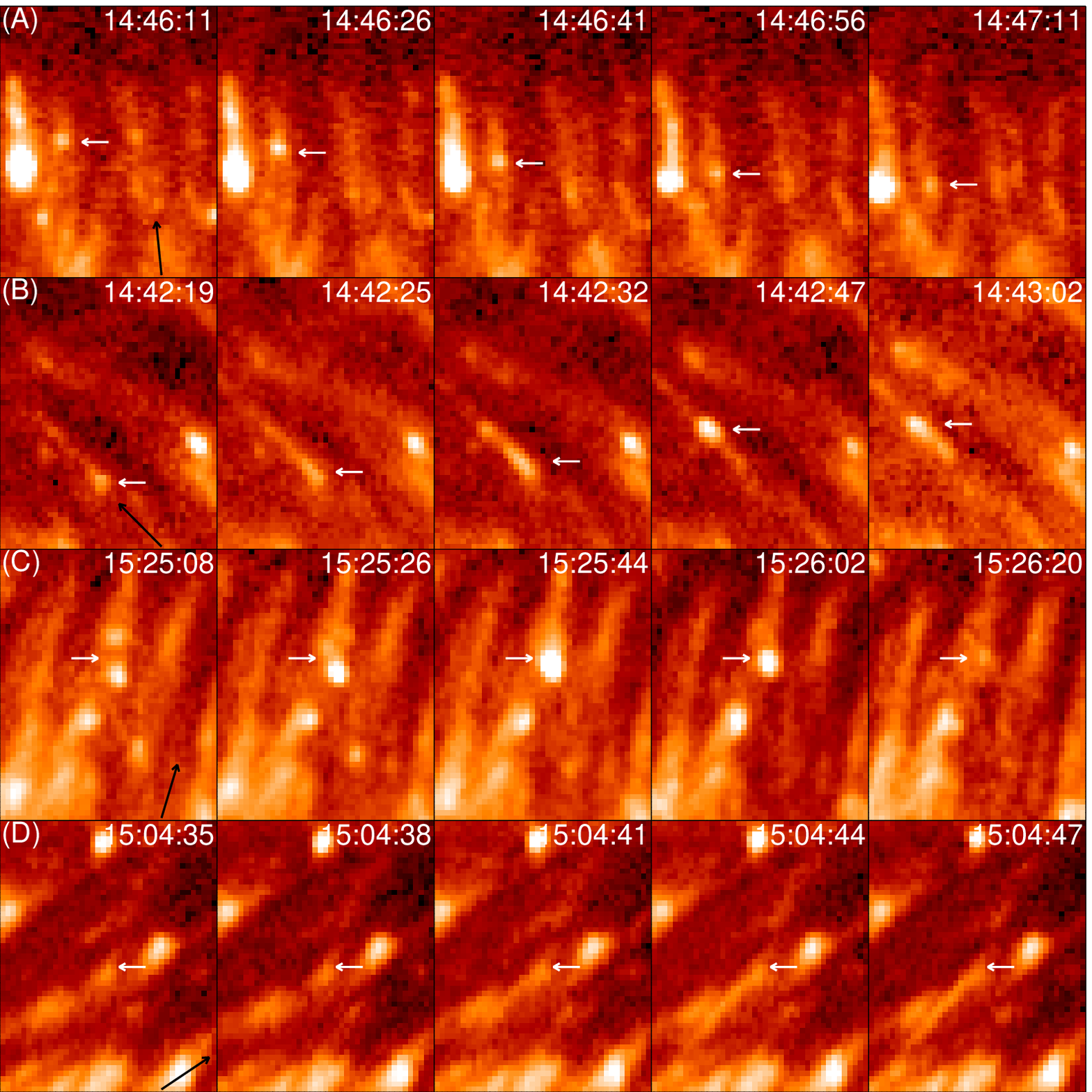}} \caption{ Temporal evolution of some penumbral bright dots (pointed by the white arrows) in IRIS 1400\AA{} images.
(A) An outward moving dot. (B) An inward moving dot. (C) Two dots merge. (D) Initiation of a fast jet from a dot. The black arrows point toward the sunspot center. The size of each image is 7$^{\prime\prime}$$\times$8$^{\prime\prime}$.} \label{fig.2}
\end{figure*}

Figure~\ref{fig.2} shows the temporal evolution of some bright dots in four different regions. The dots seem to be located in or at the edge of
the bright filamentary structures. They are often more extended in one direction. For some dots this direction
is parallel to the bright filamentary structures roughly in the sunspot radial direction. However, for other dots this direction deviates from the
radial direction. Some dots can move towards each other and eventually merge into one dot (e.g., Figure~\ref{fig.2}(C)). We often see an obvious
increase of the intensity when two dots merge. A few jet-like features with apparent speeds of $\sim$400~km~s$^{-1}$ are initiated in close proximity to some bright dots (e.g., Figure~\ref{fig.2}(D)). However, the associated bright dots seem to be present even before the jets and their intensities show no dramatic increase as the jets appear.

\begin{figure*}
\centering {\includegraphics[width=\textwidth]{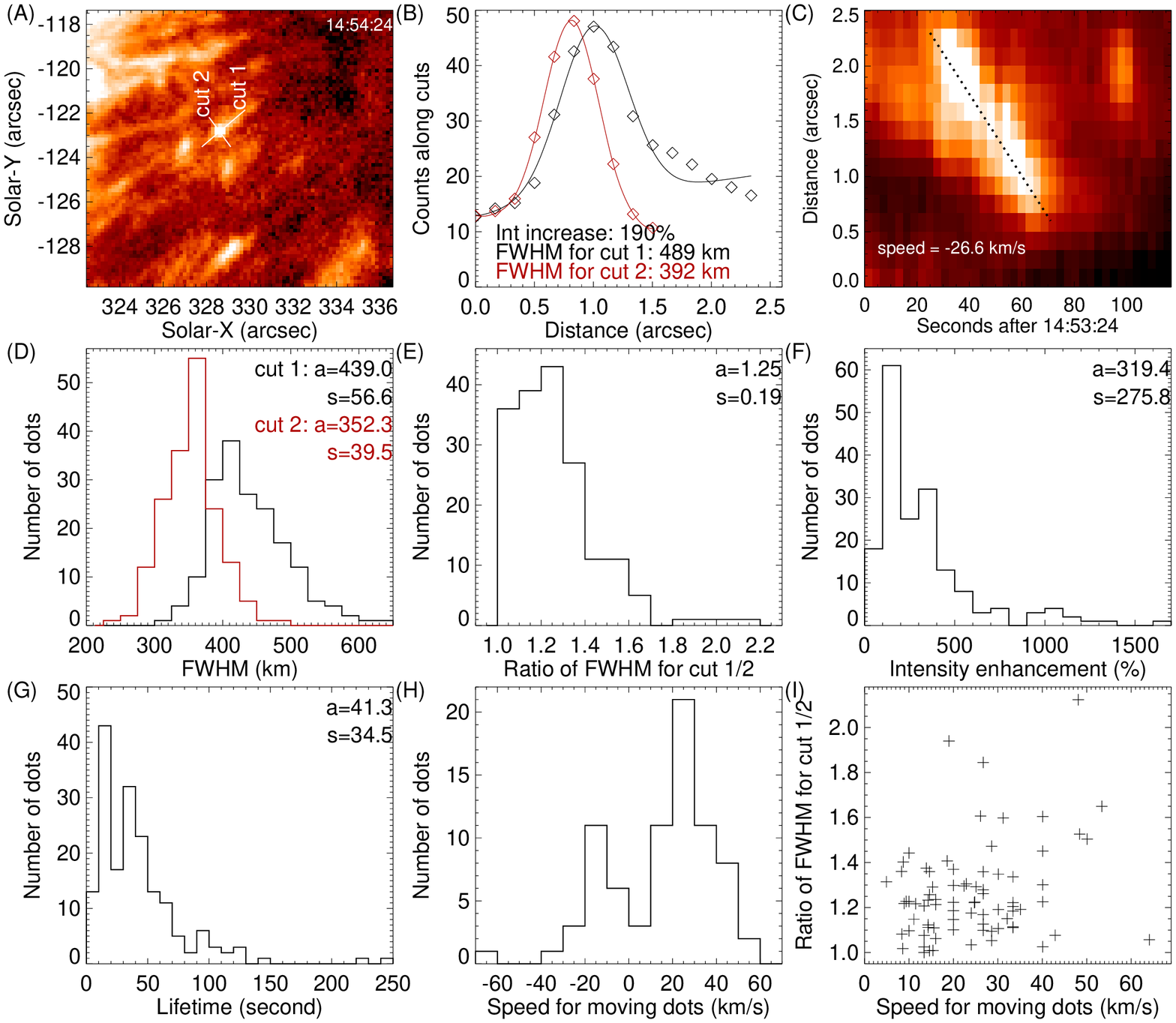}} \caption{ Quantifying the penumbral bright dots. (A) An IRIS 1400\AA{} image. Two
cuts crossing a bright dot are shown in white. (B) Intensities along the two cuts (diamonds) fitted with a Gaussian function (solid line). Results
for cuts 1 and 2 are shown in black and red, respectively. (C) Space-time plot for cut 1. (D)-(H) Distributions of the size (FWHM of the
intensity profiles along cuts 1 and 2), ratio of FWHM for cut 1/2 (length/width), intensity enhancement, lifetime and speed (positive - inward, negative - outward) for the bright dots.
Here a and s represent the average value and standard deviation, respectively. (I) Scatter plot of the relationship between the speed (absolute value) and the length/width ratio for a subset of bright dots with obvious movement.} \label{fig.3}
\end{figure*}

We have selected 176 penumbral bright dots that are well-isolated and have no obvious interaction with other dots from the 1400\AA{} images.
We first find the time when each dot shows the maximum brightness and plot the image at that time. We then plot the intensities along two cuts
across each dot: cut 1 along the long side (often radial direction) and cut 2 along the short side (often perpendicular to the radial
direction). If the dot is round, then cuts 1 and 2 are along and perpendicular to the radial direction respectively.  Each intensity profile is approximated 
with a Gaussian function plus a linear background. The full width
at half maximum (FWHM) is regarded as the size (length from cut 1 and width from cut 2) of the dot. We have also
calculated the ratio of the peak intensity and average background for each cut, and the smaller one of the two values is taken as the intensity
enhancement. We find that 44.3\% of the dots show obvious movement, generally in (or slightly deviated from) the radial direction. Using the space-time technique, their apparent speeds are estimated from the slopes of the inclined strips in space-time maps. An example of this calculation is presented in Figure~\ref{fig.3}(A)-(C).

Figure~\ref{fig.3}(D)-(H) presents the distributions of these parameters. We find that these penumbral dots are usually 300--600 km long and 250--450
km wide. The distribution of the ratio between length and width peaks at 1.25. Their lifetimes are mostly less than one minute, although some
dots last for a few minutes. Their intensities are usually a few times and sometime more than 10 times stronger than the
intensities of the surrounding filamentary structures in the slit-jaw images. Among the 78 dots which show obvious movement, 56 of them are moving
inward to the umbra and the rest 22 are moving outward. The moving dots usually have speeds of $\sim$10--40~km~s$^{-1}$. We find a weak positive correlation between the speed and the length/width ratio (Figure~\ref{fig.3}(I)).


\begin{figure*}
\centering {\includegraphics[width=\textwidth]{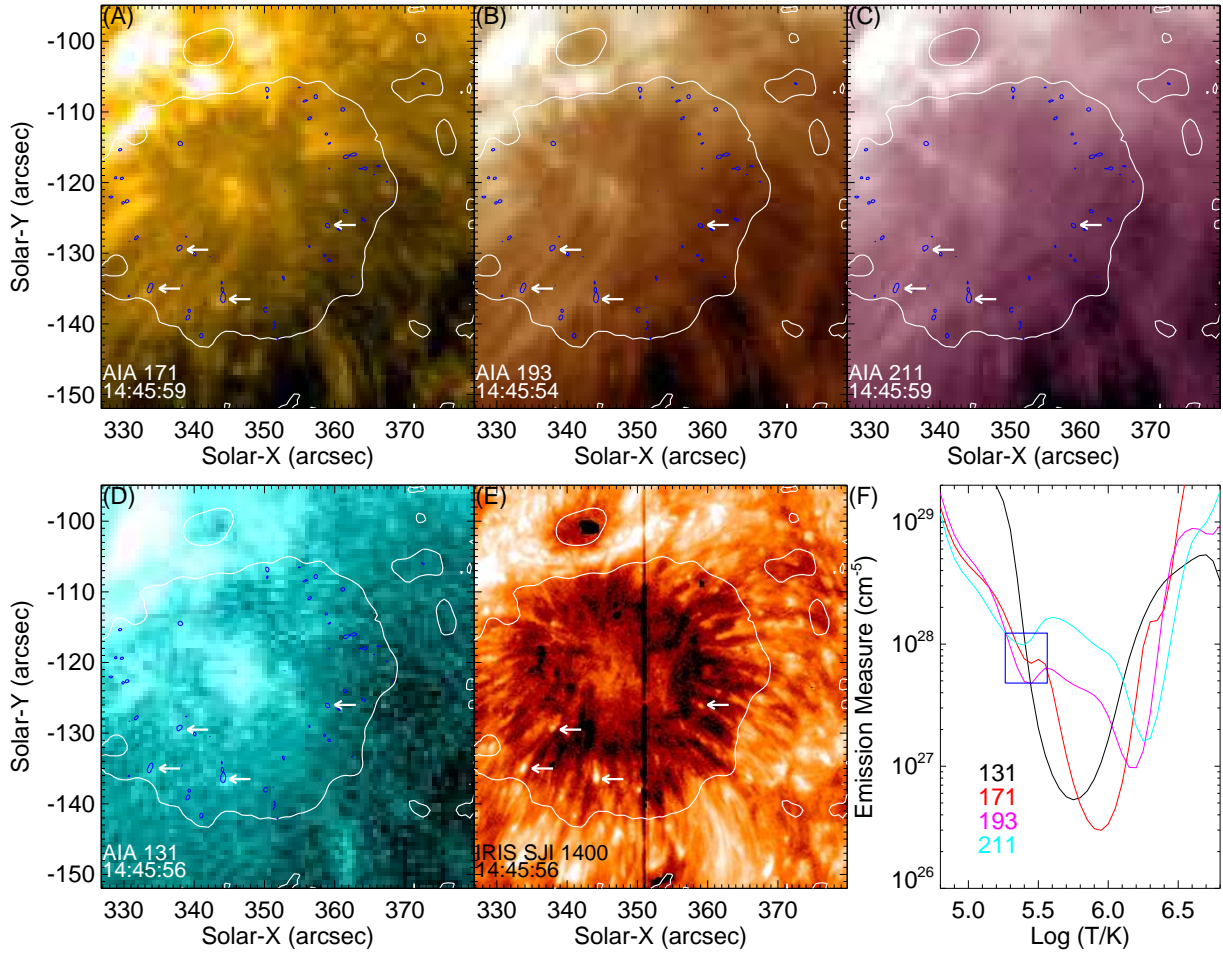}} \caption{ (A)-(E) AIA and IRIS images taken around 14:45:56 UT. Contours are the
same as in Figure~\ref{fig.1}. (F) The EM-loci curves for the bright dot at
(345$^{\prime\prime}$, -136$^{\prime\prime}$). The blue box shows a region with many crossings of the EM-loci curves. }
\label{fig.4}
\end{figure*}

Some very bright and relatively large dots also show up in AIA images, identified mainly in 304\AA{}, 171\AA{}, 193\AA{}, 211\AA{}, and 131\AA{}
passbands (e.g., the four dots pointed by the arrows in Figure~\ref{fig.4}). A few such dots last even longer than $\sim$10 min (not counted in Figure~\ref{fig.3}). They seem to be located at the footpoints of loop-like structures. Given the significant response around 10$^{5.5}$~K \citep{Martinez-Sykora2011} for all these AIA passbands, it is possible that they are cool TR structures. Determination of the temperature from differential emission measure analysis is difficult since there is little temperature discrimination at TR temperatures in the AIA channels \cite[e.g.,][]{DelZanna2011,Testa2012}.
Instead, we divide the AIA intensities by the temperature responses to calculate the EM-loci (EM: emission measure) curves \cite[e.g.,][]{DelZanna2002} to determine the possible temperature. The curves in Figure~\ref{fig.4}(F) appear somewhat consistent with an isothermal plasma of $\sim$10$^{5.4}$K  and an EM of $\sim$8$\times$10$^{27}$~cm$^{-5}$, as there are many crossings of the AIA curves in a relatively narrow box spanning 10$^{5.25}$--10$^{5.55}$K and (5 -- 12)$\times$10$^{27}$~cm$^{-5}$. However, a quantitative investigation reveals that there is no isothermal solution in the vicinity of 10$^{5.4}$K that completely fits the AIA observations. A single temperature EM model was manually fitted to AIA data for the six Fe-dominated EUV channels, varying both the temperature and EM with sufficient resolution to smoothly resolve the minimum chi-square location. The best fit reduced chi-square value is 9.3 for an isothermal model at T = 10$^{5.3}$K with EM = 1.05$\times$10$^{28}$~cm$^{-5}$, which indicates a very poor fit. Thus, we do not believe that the dot plasma is isothermal --- there must be plasma at multiple temperatures. The EM-loci curves should just be considered as an upper boundary to the true emission measure distribution.







\begin{figure*}
\centering {\includegraphics[width=\textwidth]{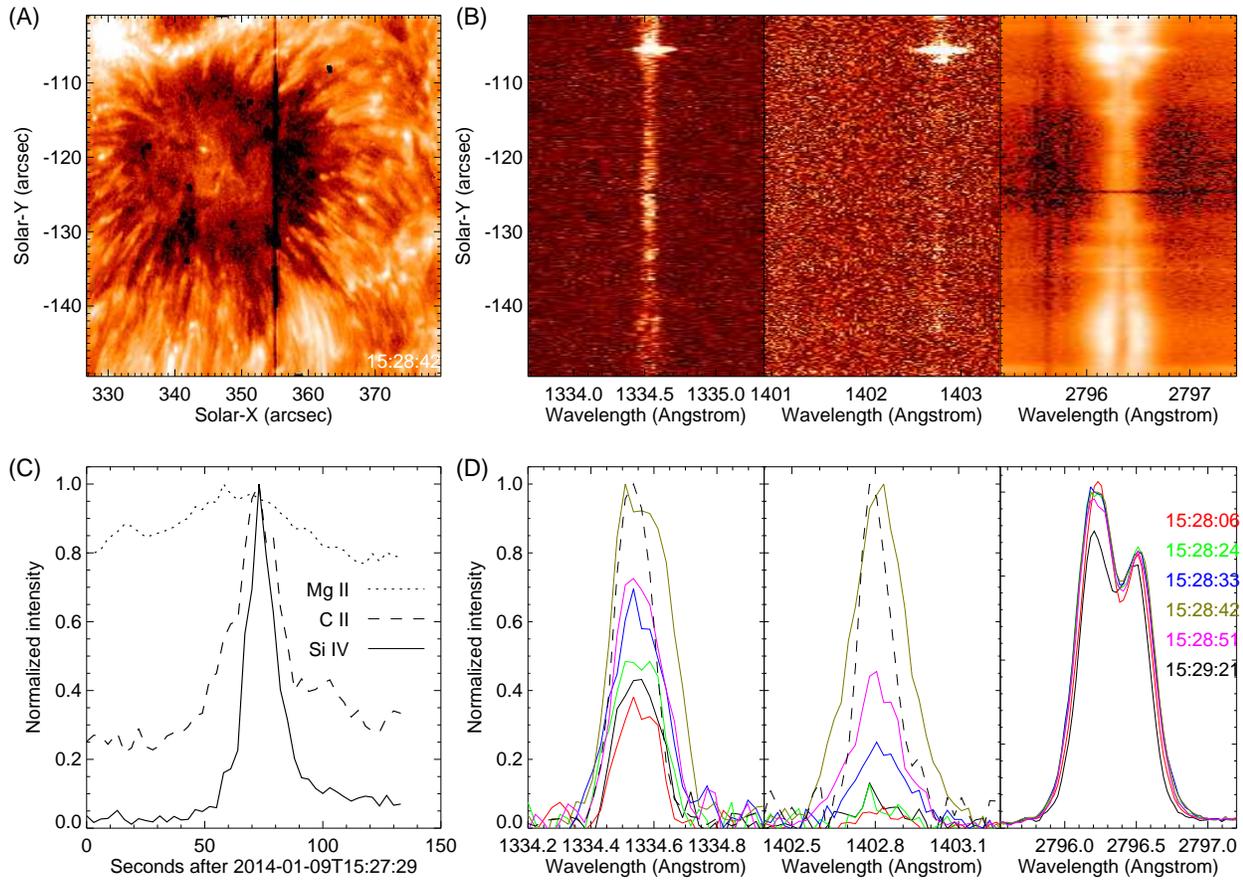}} \caption{ Spectral characteristics of a bright dot. (A) An IRIS 1400\AA{} SJI image
showing a bright dot crossing the slit. (B) Spectra of C~{\sc{ii}}~1334.53\AA{}, Si~{\sc{iv}}~1402.77\AA{}, and Mg~{\sc{ii}}~K~2796.35\AA{}
along the slit at 15:28:42. (C) Temporal evolution of the normalized line intensities. (D) Line profiles at six different times. The average line profiles (normalized) of C~{\sc{ii}} and Si~{\sc{iv}} in the penumbra are shown as the dashed lines.} \label{fig.5}
\end{figure*}

The IRIS slit crossed a few bright dots. An example is shown in Figure~\ref{fig.5}. The TR line profiles, especially the Si~{\sc{iv}} line
profiles of these bright dots, are significantly enhanced and broadened with respect to the average penumbral profiles. For the eight bright dots we
analyzed, six of them show no detectable time lag between the maximum intensities of the C~{\sc{ii}} and Si~{\sc{iv}} lines (e.g.,
Figure~\ref{fig.5}(C)). For the other two dots the Si~{\sc{iv}} intensity peak lags that of C~{\sc{ii}} by $\sim$6 seconds. Such a result seems to
favor a heating process rather than a cooling process as the formation temperatures of Si~{\sc{iv}} and C~{\sc{ii}} are 10$^{4.9}$K and
10$^{4.4}$K respectively. The Si~{\sc{iv}} line intensity is often enhanced by one to two orders of magnitude during the crossing of these bright
dots, whereas the C~{\sc{ii}} intensity is only enhanced by a factor of $\sim$3. The Mg~{\sc{ii}}~K line often shows a much weaker intensity enhancement, suggesting that some dots do not have significant chromospheric emission. Figure~\ref{fig.5}(D) shows
that the Si~{\sc{iv}} line is significantly broadened, with obvious enhancement at the wings. A small red shift ($\sim$15 km~s$^{-1}$) is found for this dot. 

We can estimate the density for the bright dot shown in Figure~\ref{fig.5}. By assuming a filling factor of one, the Si~{\sc{iv}} line intensity can be approximated as \citep{Tian2009}

\begin{equation}
\emph{$I=\frac{0.8}{4\pi}G({\it T}_{max}, {\it N}_{e}){\it N}_{e}^2l$}\label{equation1},
\end{equation}

\noindent The radiance of the Si~{\sc{iv}}~1402.77\AA{} line $I$ can be calculated using the observed counts and the effective area of IRIS \citep{DePontieu2014}. Using the coronal abundances of \cite{Schmelz2012} and the standard CHIANTI v7 ionization equilibrium \citep{Landi2012}, we can calculate the contribution function $G({\it T}_{max}, {\it N}_{e}$) at its peak temperature (${\it T}_{max}$=10$^{4.9}$K) at different densities (${\it N}_{e}$). We find a number density of $\sim$2$\times$10$^{10}$~cm$^{-3}$ if we reasonably assume an emission depth of 400~km, the typical size of the dots. 

The total thermal energy of the bright dot plasma can be calculated as

\begin{equation}
\emph{$E=3{\it N}_{e}k_{B}TV$}\label{equation1},
\end{equation}

\noindent where $k_{B}$ is the Boltzmann constant. The emission volume $V$ can be estimated as $\frac{4}{3}\pi r^3$, with $r$ being half of the typical dot size 400~km. Using a temperature ($T$) of 10$^{4.9}$K, the energy is estimated to be of the order of 10$^{22}$--10$^{23}$~erg. This is close to the typical nanoflare energy, $\sim$10$^{24}$~erg
\citep{Parker1988}.

Another characteristic of the spectra of many bright dots is the absence of the O~{\sc{iv}}~1401.16\AA{} line emission (Figure~\ref{fig.5}(B)). In principle the dominance of collisional de-excitation from the meta-stable level over radiative decay could lead to the absence of emission from this forbidden line. However, the derived density does not seem extremely high. The absence of O~{\sc{iv}}~1401.16\AA{} may also be caused by non-equilibrium ionization. The ionization and recombination times of Si~{\sc{iv}} are found to be of the order of 100 s at a much lower density \citep{Peter2006}. These time scales will be reduced and likely comparable to or smaller than the typical lifetimes of bright dots at a density of $\sim$2$\times$10$^{10}$~cm$^{-3}$. Thus, ionization equilibrium can be reached for Si~{\sc{iv}} and non-equilibrium ionization effects may not explain the disappeared/weak O~{\sc{iv}}~emission.
A recent investigation suggests that the O~{\sc{iv}}~1401.16\AA{} line
will be suppressed in the case of non-Maxwellian distribution \citep{Dudik2014}. If this is true, the absence of the O~{\sc{iv}}~1401.16\AA{} emission may
suggest that the dot plasma is far from thermal equilibrium. 

In this observation only the 1400\AA{} filter was used to achieve a very high cadence. We have examined several other short-exposure and high-cadence observations starting at 16:09 on 2013 September 1, 11:14 on 2013 December 3, 13:00 on 2014 January 9, 08:43 on 2014 February 22,  00:26 on 2014 March 16, and 01:37 on 2014 March 16. Similar bright dots are clearly present in (even beyond) the penumbrae of all sunspots. Some observations also include the 1330\AA{} channel, albeit at a lower cadence, show that the bright dots are also seen in 1330\AA{} images. Both the 1400\AA{} and 1330\AA{} filters have significant contribution from TR emission \citep{DePontieu2014}. Signatures of these dots are mostly not very prominent in the chromospheric 2796\AA{} SJI images. Bright dot-like features are also observed in light bridges and apparent oscillatory motion of them has been found in the 2013 September 1 observation.

These bright dots seem to be different from the well-known penumbral grains in photospheric images \citep{Muller1973}. Although inward and outward moving as well as static penumbral grains are all observed to exist, their moving speeds are mostly less than 1~km~s$^{-1}$. Moreover, the average lifetime of the penumbral grains is $\sim$10 minutes \citep{Zhang2013}. It is possible that some long-lasting and moveless dots we identified are related to the penumbral grains since the 1400\AA{} and 1330\AA{} passbands also have contribution from the UV continuum. However, we usually do not see visible enhancement in the continuum level from the spectra. Among tens of dot spectra we examined, only a few show obvious enhancement of the Mg~{\sc{ii}} wing (formed in the photosphere) and no visible enhancement of Si~{\sc{iv}}. The association of these bright dots with moving magnetic features \citep[e.g.,][]{Zhang2007,SainzDalda2008} is not obvious from the HMI magnetograms. However, more detailed analysis should be performed using higher-resolution magnetograms before excluding this possibility. Small-scale jets and brightenings with similar lifetimes have been found in the chromosphere above sunspots \citep{Katsukawa2007,Reardon2013}. The weak enhancement of the chromospheric Mg~{\sc{ii}}~K line emission suggests that some bright dots could be related to these chromospheric transients. 

The three-dimensional configuration of penumbral magnetic structures includes an almost horizontal field associated with penumbral filaments and a more vertical background field \citep[e.g.,][]{Solanki1993}. We find that some bright dots are located at the edge of filamentary structures and that the extension of some dots deviates from the radial direction. Also some dots are clearly located at the bases of loop-like structures, which may be considered as the background field. Perhaps some dots are generated as a result of interaction, e.g., reconnection, between the two field components. In that case the bright dots may be explained as the plasma heated to TR temperatures during reconnection. The enhanced line width might be related to reconnection outflows \citep{Innes1997}. The movement of some dots may be explained as the spreading of the heated plasma along field lines. By assuming the same velocity for all dots, both the length/width ratio and the speed projected onto the plane of sky will be larger if the field lines are more inclined. This scenario might explain the positive correlation shown in Figure~\ref{fig.3}(I). However, other probabilities such as successive reconnection in a group of crowded loops cannot be ruled out as the correlation is not very strong. 

Recently, \cite{Regnier2014} reported subarcsec bright dots from observations of the High-resolution Coronal (Hi-C) imager. These EUV dots were interpreted as signatures of impulsive energy release at the bases of trans-equatorial loops. Although not observed in sunspots, their lifetimes and sizes are similar to those of some dots we report here. We can not exclude the possibility that a similar mechanism applies for our sunspot bright dots. Highly variable brightenings in moss regions were also identified from the Hi-C data \citep{Testa2013}. However, they are located at the footpoints of bright hot loops seen in AIA
94\AA{}, which is not the case for our sunspot bright dots. So their scenario, coronal nanoflare in the overlying hot loops, may not
explain our observation. Also using the Hi-C data, \cite{Winebarger2013} reported faint, cool ($\sim$10$^{5}$K), and dense ($\sim$10$^{10}$
cm$^{-3}$) loops in an inter-moss region. These short-lived loops were interpreted as a natural consequence of impulsive energy release by
reconnection of braided magnetic fields in the TR. It would be interesting to examine whether this scenario can produce dot-like TR features.

Using IRIS observations, \cite{Kleint2014} reported bursts of high Doppler shifts implying downflows of up to 200~km~s$^{-1}$
in chromospheric and TR lines that are correlated with brightenings in slit-jaw images. These brightenings appear as dots or ribbons in
mainly the umbrae of some sunspots. The bright dots we report here are a different type of brightenings: found in mainly the penumbrae of all sunspots, usually isolated, and often moving inward or outward with speeds less than 40~km~s$^{-1}$. Despite that, it is still possible that a few of our dots are associated with falling plasma as proposed by \cite{Kleint2014}.

\section{Conclusion}
We have found many small-scale bright dots in the IRIS 1400\AA{} and 1330\AA{} slit-jaw images of sunspots. These bright dots are observed in the penumbrae of all sunspots we inspected, and are occasionally present in the umbrae and light bridges of some sunspots.

We have statistically quantified the bright dots in the penumbra of one sunspot. In 1400\AA{} slit-jaw images they are found to be generally 300--600 km long and 250--450 km wide, last mostly less than one minute, often show an intensity enhancement by a few factors, and sometimes move inward ($\sim$32\%) or outward ($\sim$13\%) with speeds of $\sim$10--40~km~s$^{-1}$. 

We have obtained spectra of several bright dots. Based on the greatly enhanced Si~{\sc{iv}}~1402.77\AA{} emission, we have roughly estimated the energy, which appears to fall in the energy range of nanoflares. The broadened line profiles of Si~{\sc{iv}}~1402.77\AA{}, together with the absence of the O~{\sc{iv}}~1401.16\AA{} line, suggests the involvement of a possible heating process.

We have discussed possible generation mechanisms for these sunspot bright dots. Many of them are likely generated by impulsive reconnection in the TR and chromosphere. Some are probably associated with falling plasma.

\begin{acknowledgements}
IRIS is a NASA small explorer mission developed and operated by LMSAL with mission operations executed at NASA Ames Research center and major
contributions to downlink communications funded by the Norwegian Space Center (NSC, Norway) through an ESA PRODEX contract. This work is
supported by contracts 8100002705 and SP02H1701R from Lockheed-Martin to SAO. We thank I. Hannah, J. Dudik, A. Lagg and L. Rouppe van der Voort for
helpful discussion.
\end{acknowledgements}

\end{document}